\documentclass[%
aps,
pre,
superscriptaddress,
tightenlines,
showpacs,showkeys,
a4paper,
12pt,
reprint,
notitlepage
]{revtex4-1}
\usepackage[english]{babel}
\usepackage{amssymb,amsmath,stmaryrd,array}

\usepackage{graphicx}
\usepackage{subfig}

\makeatletter

\newcommand{\sca}[2]{\ensuremath{\bigl({#1}\cdot{#2}\bigr)}}
\newcommand{\avr}[1]{\ensuremath{\langle{#1}\rangle}}

\newcommand{\diag}{\mathop{\rm diag}\nolimits}

\newcommand{\myarrow}[1]{\ensuremath{\xrightarrow{{#1}^\circ\mathrm{C}}}}

 \newcommand{\bs}[1]{\boldsymbol{#1}}
 \newcommand{\vc}[1]{\mathbf{#1}}
 
 \newcommand{\uvc}[1]{\hat{\mathbf{#1}}}
 
 \newcommand{\ind}[1]{\mathrm{#1}}

\newcommand{\dd}{\mathrm{d}}

\newcommand{\eff}{\mathrm{eff}}

\makeatother

\begin{document}
\DeclareGraphicsExtensions{.eps,.jpg,.png}
\title{
Orientational ``Kerr effect'' and phase modulation of light in deformed-helix ferroelectric liquid
crystals with subwavelength pitch
}

\author{Eugene~P.~Pozhidaev}
\email[Email address: ]{epozhidaev@mail.ru}
 \affiliation{%
P.N. Lebedev Physics Institute of Russian Academy of Sciences,
Leninsky prospect 53, 117924 Moscow, Russia
 }
\affiliation{%
 Hong Kong University of Science and Technology,
 Clear Water Bay, Kowloon, Hong Kong
 }

\author{Alexei~D.~Kiselev}
\email[Email address: ]{kiselev@iop.kiev.ua}
\affiliation{%
 Institute of Physics of National Academy of Sciences of Ukraine,
 prospekt Nauki 46,
 03028 Ky\"{\i}v, Ukraine}
\affiliation{%
 Hong Kong University of Science and Technology,
 Clear Water Bay, Kowloon, Hong Kong
 }

 \author{Abhishek Kumar Srivastava}
 \email[Email address: ]{Abhishek\_srivastava\_lu@yahoo.co.in}
\affiliation{%
 Hong Kong University of Science and Technology,
 Clear Water Bay, Kowloon, Hong Kong
 }

 \author{Vladimir~G.~Chigrinov}
 \email[Email address: ]{eechigr@ust.hk}
\affiliation{%
 Hong Kong University of Science and Technology,
 Clear Water Bay, Kowloon, Hong Kong
 }

 \author{Hoi-Sing~Kwok}
\email[Email address: ]{eekwok@ust.hk}
\affiliation{%
 Hong Kong University of Science and Technology,
 Clear Water Bay, Kowloon, Hong Kong
 }

\author{Maxim~V.~Minchenko}
 \affiliation{%
P.N. Lebedev Physics Institute of Russian Academy of Sciences,
Leninsky prospect 53, 117924 Moscow, Russia
 }

\date{\today}

\begin{abstract}
  We study both theoretically and experimentally the electro-optical
  properties of vertically aligned deformed helix ferroelectric liquid
crystals (VADHFLC) with subwavelength pitch
that are governed by the electrically induced optical biaxiality 
of the smectic helical structure. 
The key theoretical result is that
the principal refractive indices of homogenized VADHFLC cells
exhibit the quadratic nonlinearity
and such behavior might be interpreted as  the orientational ``Kerr effect'' caused by
the electric-field-induced orientational distortions of the FLC helix.
In our experiments, it has been observed that, 
for sufficiently  weak electric fields, the magnitude of biaxiality is proportional to the square of electric field
in good agreement with our theoretical results for the effective dielectric tensor of VADHFLCs.
Under certain conditions, the 2$\pi$ phase modulation of light, which
 is caused by one of the induced refractive indices, is observed
  without changes in ellipticity of incident light.
\end{abstract}


 \maketitle

\section{Introduction}
\label{sec:intro}

Nowadays, high-speed, low power consuming phase modulation is in high demand
for various applications such as displays, tunable gratings, beam
steering and several other photonic
devices~\cite{Engstrom:aplopt:2009,Ren:apl:2005,Martinez:aplopt:2009,Hisakado:advmat:2005}. 
Liquid crystal (LC) phase modulators are very popular for this purpose and
among all LC phases, nematics are widely used. 
However, nematics are known to have slow
response time and, in addition, this slow response gets worse 
if the LC layer thickness increases in
order to obtain a $2\pi$ phase modulation. 
Therefore, many efforts are in progress to optimize
the various LC electro-optical modes for the high speed phase
modulations. 
To improve the
response time, some recent work on faster switching LC modes has included polymer
stabilized “blue phase” (PSBPLC) mode and chiral nanostructured devices based on Kerr
effect~\cite{Ren:apl:2005,Hasebo:advmat:2005,Hisakado:advmat:2005,Yan:apl:2010,Cheng:apl:2011,Yan:apl:2013,Zhu:apl:2013}. 

The Kerr constant for the PSBPLC is generally smaller
as compared 
with other tunable LC phase 
modulators~\cite{Ren:apl:2005,Hisakado:advmat:2005}.
So, this mode is characterized by a relatively small phase change. 
There is, however, the  new class of PSBPLS 
recently reported in Refs.~\cite{Yan:apl:2010,Cheng:apl:2011,Yan:apl:2013,Zhu:apl:2013}
that reveals higher values of the Kerr constant.
For these materials,
the fabrication procedure is very complicated
as it requires
special processing condition 
such as ultraviolet curing at a precise temperature
within the LC blue phase or isotropic phase, and 
the mixture of several components~\cite{Yan:apl:2010,Cheng:apl:2011,Yan:apl:2013}.

 


Ferroelectric liquid crystals (FLCs) 
representing another and most promising candidate 
are characterized by very fast response time.
However, most of the FLC modes are not suitable for pure phase
modulation devices because their optical axis sweeps in 
the plane of the cell substrate
producing changes in the polarization state of the incident light. 
In order to get around
the optical axis switching problem
the system consisting of a FLC half-wave plate sandwiched between two 
quarter-wave plates was suggested in Ref.~\cite{Love:optcom:1994}.
But, in this system, the $2\pi$ phase modulation requires the smectic
tilt angle to be equal to 45 degrees. 
This value is a real challenge for the material science. 
Even though some of the
antiferroelectric and ferroelectric liquid
crystals possess the desired tilt angle, the difficulty is that 
the response time dramatically increases when the tilt angle
grows up to 45 degrees~\cite{Barnik:mclc:1987,Pozhidaev:jetp:1988}.

In this article, we deal with the approach to phase modulation 
that uses the electro-optical properties of
helical structures in deformed helix ferroelectric liquid crystals 
(DHFLC)~\cite{Beresnev:lc:1989,Chigr:1999,Lee:optex:2005,Kiselev:pre:2011}.
Our goal is to systematically examine 
the physical characteristics that govern
phase modulation 
in
vertically aligned DHFLC (VADHFLC) cells where 
the twisting axis of FLC helix is normal to the substrates
and the helix pitch, $P$, is short as compared 
to the wavelength of light, $\lambda$.

The paper is organized as follows.
 
In Sec.~\ref{sec:theory}, we
begin with the theoretical analysis which is based on
a generalized version of
the theory developed in Ref.~\cite{Kiselev:pre:2011}.
For the effective dielectric tensor
of homogenized VADHFLCs with the subwavelength pitch,
it turned out that, 
in agreement with the results of the
experimental study~\cite{Lee:optex:2005}
where the VADHFLC mode was used
to achieve the $2\pi$ phase modulation at the wavelength 
1.55~$\mu$m,
in-plane rotations of the optical axes are suppressed.
Optical anisotropy of short-pitch VADHFLCs is 
found to be generally biaxial and,
in the low voltage regime,
the electric field dependence of the refractive indices
exhibits the quadratic nonlinearity
that can be
interpreted as the \textit{orientational ``Kerr effect''}.

Electro-optical properties of VADHFLC cells with the subwavelength
helix pich are studied experimentally in Sec.~\ref{sec:experiment}.  
Experimental details are given in Sec.~\ref{subsec:setup},
where we describe the samples and the setup
employed to perform measurements.
The experimental results
verifying the theoretical predictions
are presented in Sec.~\ref{subsec:results}.

Finally, in Sec.~\ref{sec:conclusion},
we draw the results together and make some concluding remarks.

\begin{figure}[!tbh]
  \includegraphics[width=8cm]{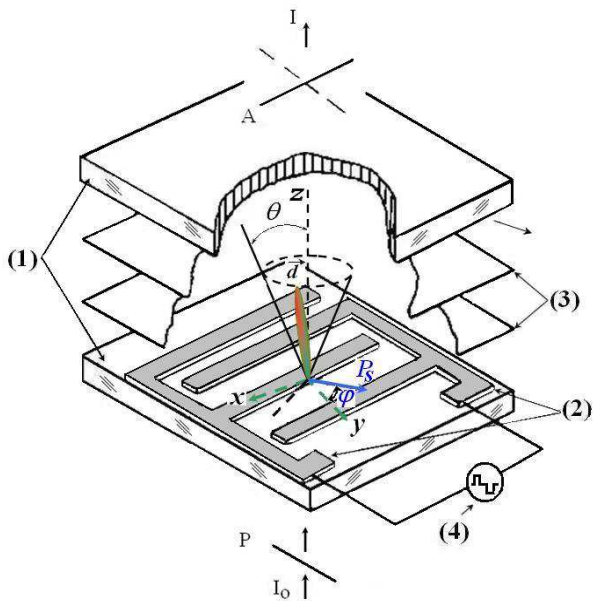}
  \caption{%
(Color online)
    Schematic representation of geometry of a vertically aligned DHFLC
    cell: (1)~are glassy plates; (2)~are in-plane electrodes; (3)~are
    smectic layers parallel to the glassy substrates; (4)~is a voltage
    generator; P (A) is polarizer (analyzer); $\theta$ is the tilt angle of
    director $\vc{d}$ in smectic layers; $\varphi$ is the azimuthal angle of the
    spontaneous polarization vector $\vc{P}_s$; $I_0$ is intensity of incident
    light; $I$ is intensity of light transmitted through the VADHFLC
    cell placed between P and A.}
\label{fig:geom}
\end{figure}

\section{Theory}
\label{sec:theory}

Optical properties of planar aligned DHFLC cells 
with subwavelength
pitch, where the axis of FLC helix, $\uvc{h}$,
is parallel to the substrates,
was described in terms of
the effective dielectric tensor of 
homogenized FLC helical structure in Ref.~\cite{Kiselev:pre:2011}.
The general result of Ref.~\cite{Kiselev:pre:2011} 
is that under the action of 
the electric field applied along the normal to the twisting axis,
$\vc{E}\perp \uvc{h}$,
a DHFLC cell, in which the zero-field optical anisotropy
is uniaxial with the optical axis parallel to the helical axis
$\uvc{h}$, 
becomes biaxially anisotropic with the principal optical axes
rotated about the vector of applied electric field, $\vc{E}$.

\begin{figure}[!tbh]
\includegraphics[width=9cm]{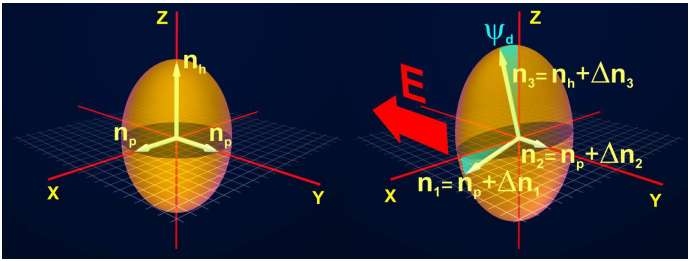}
\caption{%
(Color online)
Ellipsoids of effective refractive indices of a short-pitch
VADHFLC cell.
Left:~in the case of zero electric field, $E=0$~V, 
the effective dielectric tensor~\eqref{eq:eps_eff_zero} is uniaxially anisotropic 
with 
the optical axis parallel to the helix axis,
$\uvc{h}=\uvc{z}$.  
Right:~
applying an in-plane electric field,
$\vc{E}$,
makes
the optical anisotropy biaxial
$n_1\ne n_2 \ne n_3$
with the two optical axes rotated by the angle
$\psi_d$ 
(see equation~\eqref{eq:eigenvector})
about the electric field vector parallel to 
the third optical axis.
}
\label{fig:ellipsoids}
\end{figure}

In this section our task is to show that 
similar behaviour applies 
to the case of the slab geometry
of vertically aligned DHFLC cells
which is depicted in Fig.~\ref{fig:geom}.
For this geometry,
the above effect is  illustrated in Fig.~\ref{fig:ellipsoids}.

To this end, we introduce
the effective  dielectric tensor,
$\bs{\varepsilon}_{\eff}$,
describing a homogenized DHFLC helical structure
characterized by the director
\begin{align}
&
\uvc{d}=
\cos\theta\,\uvc{h}+
\sin\theta\,\uvc{c},
\label{eq:director}
  \end{align}
where $\uvc{c}\perp\uvc{h}$ is the $c$-director,
that lies on the smectic cone with
the smectic tilt angle $\theta$ and rotates
in a helical fashion about a uniform twisting axis
$\uvc{h}$ forming the FLC helix with 
the subwavelength pitch, $P\ll\lambda$.
The polarization unit vector
\begin{align}
&
\uvc{p}=
\uvc{h}\times\uvc{c}\parallel \vc{P}_s,
\label{eq:pol-vector}
  \end{align}
is directed along the vector of spontaneous ferroelectric
polarization, $\vc{P}_s=P_s\uvc{p}$,
and, for a vertically aligned FLC helix
in the slab geometry shown in Fig.~\ref{fig:geom},
we have
\begin{align}
&
\uvc{h}=\uvc{z},
\quad
\uvc{c}=\cos\varphi\,\uvc{x}+
\sin\varphi\,\uvc{y},
\notag
\\
&
\uvc{p}=\cos\varphi\,\uvc{y}-
\sin\varphi\,\uvc{x},
\quad
\vc{E}=E\,\uvc{y},
\label{eq:director-vert}
 \end{align}
where $\varphi$ is the azimuthal angle
around the cone
\begin{align}
&
\varphi\approx \phi_0 +\alpha_{E}\sin\phi_0,
\quad
\phi_0=2 \pi \sca{\uvc{h}}{\vc{r}}/P=2 \pi z/P,
\label{eq:Phi}    
 \end{align}
that depends on
the applied electric field, $E$,
through the electric field parameter
$\alpha_E=\gamma_E E$ linearly proportional to
the ratio of the applied and critical electric
fields~\cite{Chigr:1999,Hedge:lc:2008}: $E/E_c$.

For a biaxial FLC, the components of the dielectric tensor,
$\bs{\varepsilon}$,
are given by
\begin{align}
&
\epsilon_{i j}=
  \epsilon_{0}
\delta_{i j}+
(\epsilon_{1}-\epsilon_{0})\,
d_i d_j
+
(\epsilon_{2}-\epsilon_{0})\,
p_i p_j=
\notag
\\
&
=
  \epsilon_{0}
(
\delta_{i j}+u_1 d_i d_j
+u_2 p_i p_j
),
\label{eq:diel-tensor}
\end{align}
where 
$i,j\in\{x,y,z\}$
and
$u_i=(\epsilon_{i}-\epsilon_{0})/\epsilon_{0}$
are the \textit{anisotropy parameters}.
Note that, in the case of uniaxial anisotropy with $u_2=0$,
the principal values of the dielectric tensor are:
$\epsilon_{0}=\epsilon_2=\epsilon_{\perp}$ and
$\epsilon_{1}=\epsilon_{\parallel}$,
where $n_{\perp}=\sqrt{\mu\epsilon_{\perp}}$
($n_{\parallel}=\sqrt{\mu\epsilon_{\parallel}}$)
is the ordinary (extraordinary) refractive index
and $\mu$ is the magnetic permittivity. 

According to Ref.~\cite{Kiselev:pre:2011},
normally incident light feels 
effective in-plane anisotropy
described by the averaged tensor,
$\avr{\bs{\varepsilon}_P}$:
\begin{align}
&
\avr{\epsilon_{\alpha\beta}^{(P)}}=
\left\langle
 \epsilon_{\alpha\,\beta}-\frac{ \epsilon_{\alpha\,z}
   \epsilon_{z\,\beta}}{ \epsilon_{zz}}
\right\rangle
=
\notag
\\
&
=
\epsilon_{0}
\left\langle
\delta_{\alpha\beta}+\frac{u_1 d_\alpha d_\beta+u_2 p_\alpha
  p_\beta+u_1 u_2 q_\alpha q_\beta}{1+u_1  d_z^2+u_2  p_z^2}\,
\right\rangle,
\label{eq:in-plane}
\\
&
q_{\alpha}=p_z d_{\alpha}-d_z p_{\alpha},
\quad 
\alpha,\beta\in\{x,y\},
\label{eq:q-vector}
\end{align}
where $\avr{\ldots}=(2\pi)^{-1}\int_{0}^{2\pi}\ldots\dd\phi_0$,
and the effective dielectric tensor
\begin{align}
  \label{eq:eff-diel-tensor}
&
\bs{\varepsilon}_{\eff}=
\begin{pmatrix}
  \epsilon_{xx}^{(\eff)} & \epsilon_{xy}^{(\eff)}& \epsilon_{xz}^{(\eff)}\\
\epsilon_{yx}^{(\eff)}& \epsilon_{yy}^{(\eff)} & \epsilon_{yz}^{(\eff)}\\
\epsilon_{zx}^{(\eff)} & \epsilon_{zy}^{(\eff)} & \epsilon_{zz}^{(\eff)}
\end{pmatrix}  
\end{align}
can be expressed in terms of the averages
    \begin{align}
&
   \eta_{zz}=
\avr{\epsilon_{zz}^{-1}}
=
\epsilon_{0}^{-1}
\avr{[1+u_1 d_z^2+u_2 p_z^2]^{-1}},
\label{eq:eta}
\\
&
    \beta_{z\alpha}=
\avr{\epsilon_{z\alpha}/\epsilon_{zz}}
=
\left\langle
\frac{u_1 d_z d_\alpha+u_2 p_z p_\alpha}{1+u_1 d_z^2+u_2 p_z^2}
\right\rangle,
\label{eq:beta}
  \end{align}
as follows
\begin{align}
&
\epsilon_{zz}^{(\eff)}
=1/\eta_{zz},
\quad
\epsilon_{z\alpha}^{(\eff)}
=\beta_{z\alpha}/\eta_{zz},
\notag
\\
&
\epsilon_{\alpha\beta}^{(\eff)}
=\avr{\epsilon_{\alpha\beta}^{(P)}}+
\beta_{z\alpha} \beta_{z\beta}/\eta_{zz}.
\label{eq:elements-eff-diel-tensor}
  \end{align}
 
General
formulas~\eqref{eq:in-plane}-\eqref{eq:elements-eff-diel-tensor}
give  the zero-order approximation
for homogeneous models
describing the optical properties of
short pitch DHFLCs.
Assuming that
the pitch-to-wavelength ratio $P/\lambda$
is sufficiently small,
these formulas 
can now be used to derive the effective dielectric tensor of
homogenized short-pitch DHFLC cell for both vertically and planar aligned
FLC helix. 
In the latter case
that was studied in Ref.~\cite{Kiselev:pre:2011}
for uniaxial FLCs, 
the relations~\eqref{eq:director-vert}
are changed as follows
\begin{align}
&
\uvc{h}=\uvc{x},
\quad
\uvc{c}=\cos\varphi\,\uvc{y}+
\sin\varphi\,\uvc{z},
\notag
\\
&
\uvc{p}=\cos\varphi\,\uvc{z}-
\sin\varphi\,\uvc{y},
\quad
\vc{E}=E\,\uvc{z}.
\label{eq:director-planar}    
  \end{align}

In this paper we concentrate on 
the zero-order homogeneous model for
vertically aligned DHFLC
cells.
From the results of Ref.~\cite{Oldano:pre:2003a}
it can be inferred that,
despite some difficulties related to 
the effect of optical rotation~\cite{Oldano:prb:1996,Oldano:pre:1998,Ponti:lc:2001},
the homogeneous models work well for many
optical properties of vertically aligned short-pitch 
FLCs. 
Note that our general approach~\cite{Kiselev:pre:2011}
can be regarded as a modified version of 
the transfer matrix method~\cite{Markos:bk:2008,Kis:jpcm:2007,Kiselev:pra:2008} and 
the perturbative technique
suggested in Ref.~\cite{Oldano:prb:1996}
can be applied
to go beyond the zero-order approximation.
An alternative approach is based on
the Bloch wave method~\cite{Galatola:pre:1997}
and has also been used to define the homogeneous models
for periodically modulated anisotropic
media~\cite{Ponti:lc:2001,Oldano:pre:2001,Etxeb:pre:2001}.

From Eq.~\eqref{eq:director-vert}, the $z$ components of
of the director, $d_z=\cos\theta$, and the polarization vector,
$p_z=0$, do not depend on $\varphi$. So, we have
\begin{align}
  \label{eq:e_zz-vert}
&
  \frac{\epsilon_{zz}}{\epsilon_{0}}\equiv
e_{zz}=1+u_1 d_z^2=
r_1\cos^2\theta+\sin^2\theta
=  \frac{\epsilon_{zz}^{(\eff)}}{\epsilon_{0}},
\\
&
\label{eq:e_za-vert}
\begin{pmatrix}
  \epsilon_{z x}^{(\eff)}/\epsilon_{0}\\
  \epsilon_{z y}^{(\eff)}/\epsilon_{0}
\end{pmatrix}
=
u_1 \cos\theta\sin\theta
\begin{pmatrix}
  \avr{\cos\varphi}\\
  \avr{\sin\varphi}
\end{pmatrix},
\\
&
\label{eq:e_ab-vert}
\epsilon_{\alpha\beta}^{(\eff)}
=\avr{\epsilon_{\alpha\beta}^{(P)}}+
\epsilon_{z \alpha}^{(\eff)} \epsilon_{z
  \beta}^{(\eff)}/\epsilon_{zz}^{(\eff)},
\\
&
2 \avr{\bs{\varepsilon}_P}/\epsilon_{0}=
(r_1/e_{zz}+r_2)
\begin{pmatrix}
  1 & 0\\
  0 & 1
\end{pmatrix}
+
\notag
\\
&
+
(r_1/e_{zz}-r_2)
\begin{pmatrix}
  \avr{\cos 2\varphi} & \avr{\sin 2 \varphi}\\
  \avr{\sin 2 \varphi} & -\avr{\cos 2\varphi}
\end{pmatrix}
\label{eq:e_p-vert}
\end{align}
where $r_i=\epsilon_i/\epsilon_0$.
After substituting the expression for the azimuthal angle~\eqref{eq:Phi}
into Eqs.~\eqref{eq:e_zz-vert}-~\eqref{eq:e_p-vert},
it is not difficult to perform averaging over the helix pitch
giving the result that
$\avr{\sin \varphi}=\avr{\sin 2 \varphi}=0$
and
\begin{align}
 &
\avr{\cos 2 \varphi}=J_2(\alpha_E)\approx
\alpha_E^2/2,
\notag
\\
&
  \avr{\cos\varphi}=-J_1(\alpha_E)\approx
-\alpha_E/2=\chi_E E/P_s,
  \label{eq:averages-vert} 
\end{align}
where $J_n(x)$ is the Bessel function of the first kind of $n$th
order~\cite{Abr}, $\chi_E=\partial \avr{P_y}/\partial E$ is
the dielectric susceptibility of the Goldstone mode~\cite{Urbanc:ferro:1991} 
and $P_y=P_s\cos\varphi$ is the $y$ component
of the polarization vector, $\vc{P}_s$.
From Eqs.~\eqref{eq:e_zz-vert}-~\eqref{eq:averages-vert},
it is clear that, in the zero-field limit with $E=0$ ($\alpha_E=0$),
the effective dielectric tensor 
\begin{align}
&
  \label{eq:eps_eff_zero}
  \avr{\bs{\varepsilon}_{\eff}}|_{E=0}=
\diag(\epsilon_p,\epsilon_p,\epsilon_h)=
\diag(n_p^2,n_p^2,n_h^2),
\\
&
  \label{eq:eps_z_zero}
\epsilon_h=\epsilon_{zz}^{(\eff)}=
\epsilon_{1}\cos^2\theta+\epsilon_{0} \sin^2\theta,
\\
&
\epsilon_p=(\epsilon_{0}\epsilon_{1}/\epsilon_{z}+\epsilon_{2})/2,
  \label{eq:eps_p_zero}
\end{align}
where $n_p$ and $n_h$
are the principal values of the refractive indices,
 is uniaxially anisotropic with the
optical axis directed along the twisting axis ($z$ axis).

Interestingly,
at $\epsilon_h=\epsilon_p$, 
the dielectric tensor~\eqref{eq:eps_eff_zero} 
becomes isotropic.
The condition of zero-field isotropy
\begin{align}
  \label{eq:zero_iso}
  r_2=2 e_{zz}-r_1/e_{zz}
\end{align}
gives the ratio $r_2\equiv \epsilon_2/\epsilon_0$
as a function of $r_1\equiv \epsilon_1/\epsilon_0$
and the tilt angle $\theta$.
According to the condition~\eqref{eq:zero_iso}, 
the ratio $r_2$ monotonically changes from
$2 r_1-1$ to $2-r_1$ as the tilt angle varies from zero
to $\pi/2$. When $n_{\perp}\approx 1.5$
and $n_{\parallel}\approx 1.7$, the ratio $r_1$
can be estimated at about $1.284$
and equation~\eqref{eq:zero_iso}
gives $r_2\approx 1.37$ at $\theta=30$~deg.
In this case, $r_2$ reaches unity
and $r_1$ at $\theta\approx 55.8$~deg
and $\theta\approx 36.3$~deg, respectively.

In the presence of the applied electric field,
$E\ne 0$, the in-plane effective dielectric tensor~\eqref{eq:e_p-vert}
is no longer isotropic
and can be written in the following form:
\begin{align}
  \label{eq:eps_p-vert-E}
  \avr{\bs{\varepsilon}_P}
=\diag(\epsilon_{+},\epsilon_{-}),
\quad
\epsilon_{\pm}
\approx
\epsilon_p
\pm
(\epsilon_p-\epsilon_2)\gamma_E^2 E^2/2.
\end{align}
So, for normally incident light, the in-plane
optical axis with the principal value of the dielectric 
tensor equal to $\epsilon_{+}$ ($\epsilon_{-}$)
is normal (parallel) to the electric field $\vc{E}$.
The refractive index difference  
\begin{align}
&
  \label{eq:detla_n-vert}
  \delta n_{\ind{ind}}=n_{+}-n_{-}\equiv
n_{\perp E}-n_{\parallel E}
\approx K_{\ind{kerr}} E^2,
\\
&
K_{\ind{kerr}}=2 n_p (1-\epsilon_2/\epsilon_p) (\chi_E/P_s)^2,
 \label{eq:detla_n-vert-kerr}
\end{align}
where $n_p=\sqrt{\mu \epsilon_p}$ is the in-plane zero-field refractive
index,
will be referred to as the 
\textit{electrically induced biaxiality (in-plane birefringence)}
and
exhibits the Kerr-like nonlinearity: $\delta n_{\ind{ind}}\propto E^2$.

For the phase modulation of normally incident light, 
it is important that orientation of the in-plane optical axes does not depend
on the applied electric field $E$. 
By contrast, propagation of the obliquely incident light
is governed by the effective dielectric tensor
\begin{align}
&
    \bs{\varepsilon}_{\eff}\approx
\begin{pmatrix}
  \epsilon_{+} + \gamma_{xz}^2 E^2/\epsilon_z& 0& \gamma_{xz} E\\
0 & \epsilon_{-} & 0\\
\gamma_{xz} E & 0 & \epsilon_{z}
\end{pmatrix},
\notag
\\
&
\gamma_{xz}= (\epsilon_1-\epsilon_0)
\cos\theta\sin\theta\chi_E/P_s,
\label{eq:eff-diel-vert}
\\
&
\epsilon_{ij}^{(\eff)}=\epsilon_{-}^{(\eff)}\delta_{ij}+
(\epsilon_{+}^{(\eff)}-\epsilon_{-}^{(\eff)}) d_i^{(\eff)}
d_j^{(\eff)}
\notag
\\
&
+
(\epsilon_{y}^{(\eff)}-\epsilon_{-}^{(\eff)})\delta_{i y} \delta_{y j}
\label{eq:diag-eff-diel-vert}
\end{align}
which is generally biaxial. From Eq.~\eqref{eq:diag-eff-diel-vert},
this  tensor is characterized by the three different
principal values (eigenvalues)
\begin{align}
&
  \label{eq:eigenvalue-y}
  \epsilon_{y}^{(\eff)}=\epsilon_{yy}^{(\eff)}=\epsilon_{-}=n_2^2,
\\
&
  \epsilon_{\pm}^{(\eff)}=(\epsilon_{xx}^{(\eff)}+\epsilon_{zz}^{(\eff)})/2
\pm \sqrt{[\Delta\epsilon]^2+[\gamma_{xz} E]^2}=n_{1,\,3}^2,
  \label{eq:eigenvalue-zy}
\end{align}
where
$\Delta\epsilon=(\epsilon_{xx}^{(\eff)}-\epsilon_{zz}^{(\eff)})/2$,
and the optical axis
\begin{align}
  \label{eq:eigenvector}
  \uvc{d}_{\eff}=\cos\psi_d\uvc{x}+
\sin\psi_d\uvc{z},
\quad
\tan(2\psi_d)=\gamma_{xz} E/\Delta\epsilon
\end{align}
is defined as the eigenvector corresponding to the principal value
$\epsilon_{+}^{(\eff)}$, whereas 
the eigenvector
for the eigenvalue $\epsilon_{y}^{(\eff)}=\epsilon_{-}$ given
in Eq.~\eqref{eq:eps_p-vert-E}
(the unit vector $\uvc{y}$) 
is parallel to the applied field $\vc{E}=E\uvc{y}$.

Clearly, 
the applied electric field 
changes the principal values of the dielectric tensor
(see Eqs.~\eqref{eq:eigenvalue-y} and~\eqref{eq:eigenvalue-zy})
and, similar to the case of planar oriented FLC helix,
it additionally results in the rotation of
the optical axes about its direction (the $y$ axis) 
by the tilt angle $\psi_d$ defined in Eq.~\eqref{eq:eigenvector}.
Behavior of the electrically induced part of the principal values
in the low electric field region is typically dominated by the
Kerr-like nonlinear terms proportional to $E^2$, whereas 
the electric field dependence of the angle $\psi_d$ is
approximately linear: $\psi_d\propto E$.

For a medium that lacks inversion symmetry,
such behavior looks unusual because, in such media,
the Kerr-like nonlinearity
is typically masked by the much stronger 
Pockels electro-optic effect~\cite{Melnichuk:pra:2010}.
Actually, it differs from
the well-known Kerr effect which is  a quadratic electro-optic effect
related to the electrically induced birefringence
in optically isotropic (and transparent) materials
and
which is mainly caused by the electric-field-induced orientation
of polar molecules~\cite{Weinberger:pml:2008}.

By contrast, in our case, similar to PSBPLCs, we deal with the effective dielectric tensor
of a nanostructured chiral smectic liquid crystal. 
The expression for the components of this tensor~\eqref{eq:elements-eff-diel-tensor}
involves averaging over the FLC orientational stucture.
Owing to the symmetry of undisorted FLC helix,
the zero-field optical anisotropy of DHFLC is generally uniaxial
with the optical axis parallel to the twisting axis.

Another effect is that, for
the orientational distortions of the FLC helix 
that are linearly dependent
on the electric field and defined in Eq.~\eqref{eq:Phi}, 
the averaging procedure yields 
the effective dielectric tensor
with diagonal (non-diagonal) elements
being an even (odd) function of the electric field.
The latter is behind 
the quadratic  nonlinearity of 
the principal values of the dielectric tensor and
the corresponding refractive indices.
Clearly, this effect is caused by
the electrically induced distortions of the helical
structure and bears some resemblance to the electro-optic Kerr effect. 
Hence it will be referred to as the orientational
``Kerr effect''.

Interestingly, the above discussed Kerr-like regime may break down 
when the condition of zero-field isotropy~\eqref{eq:zero_iso}
is fulfilled and $\epsilon_p=\epsilon_z$.
This is the special case where
$\Delta\epsilon$ is proportional to $E^2$ and
$\tan(2\psi_d)\propto 1/E$,
so that the optical axis tilt angle $\psi_d$ decreases with the electric field
starting from the zero-field value equal to $\pi/4$.
In addition, it turned out that, for low electric fields, 
the electric field dependence of
the dominating contributions to the electrically controlled
part of the principal values is linear.
 
We conclude this section with the remark
that the difference of refractive indices~\eqref{eq:detla_n-vert},
though formally defined as an in-plane birefringence
for a normally incident light,
results from the electric field induced optical biaxiality
combined with the above discussed out-of-plane rotation
of the optical axes.
This is why this important parameter is called
the electrically induced biaxiality.
In the subsequent section, the relation~\eqref{eq:detla_n-vert}
will be essential for making a comparison
between the theory and the experimental data.

\begin{figure}[!tbh]
\subfloat[]{%
\resizebox{30mm}{!}{\includegraphics{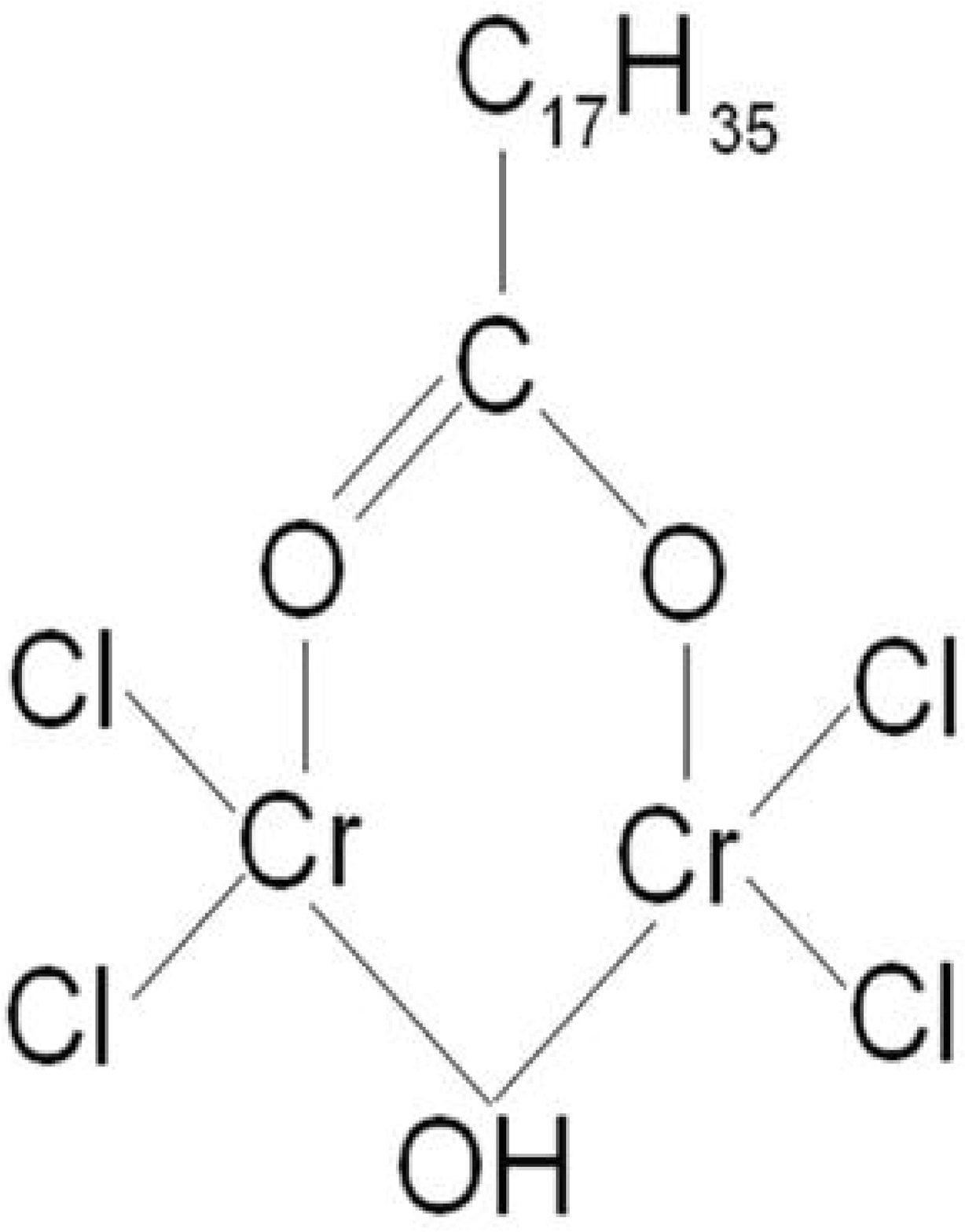}}
\label{subfig:chem}
}
\quad
\subfloat[]{%
\resizebox{45mm}{!}{\includegraphics{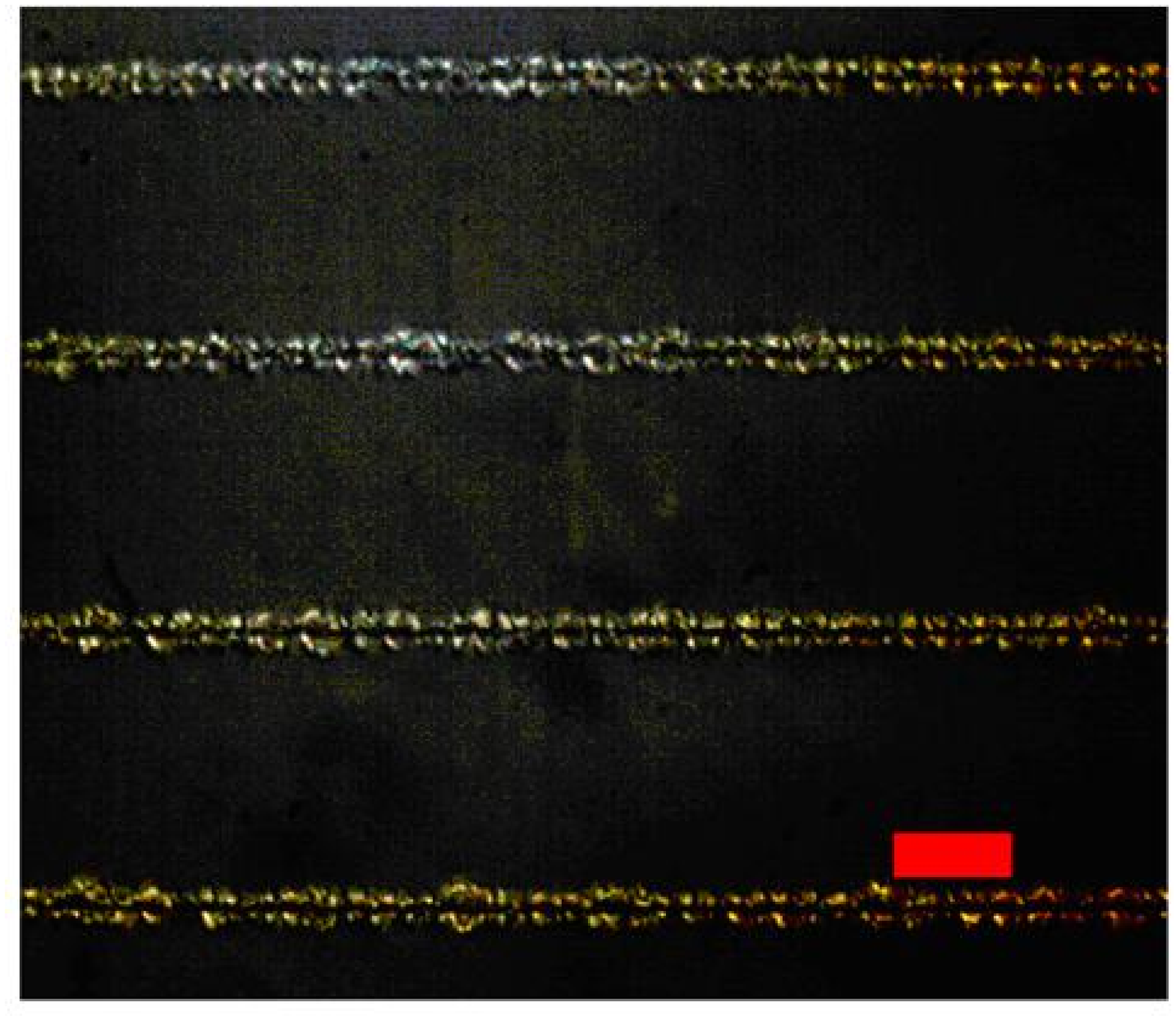}}
\label{subfig:micro}
}
\caption{%
(Color online)
(a)~Chemical structure of stearic acid
chromium salt.
(b)~Microphotograph of VADHFLC
cell placed between crossed polarizers
at $E=0$~V.
The length of red mark is 20~$\mu$m.  
}
\label{fig:chem}
\end{figure}

\section{Experiment}
\label{sec:experiment}

\subsection{Samples and experimental setup}
\label{subsec:setup}

Now we pass on to describing our experimental procedure
employed to measure the characteristics of
EO response of VADHFLC cells.

The DHFLC with smectic layers parallel to the surfaces, and with the
well-known geometry of in-plane electrodes have been used in our
experiments. Figure~\ref{fig:geom} presents an extensive overview of the
electro-optical cell with vertically aligned FLC helix and inter-digital
electrode deployed on one of the two glass plates.  The inter-digital
electrodes with electrode width 2~$\mu$m and electrode gap of
50~$\mu$m have been deployed to apply electric field parallel to the smectic
layers. The vertical alignment to the DHFLC has been accomplished by
spin coating glass plates with 40~nm thick layer of stearic acid
chromium salt (its chemical structure is shown in Fig.~\ref{subfig:chem}) 
followed by soft backing at temperature 100$^\circ$C for 10 min.
Figure~\ref{subfig:micro} represents the optical microphotograph of the VADHFLC
cell which is placed between crossed polarizers in the absence of the electric field.

The FLC-587 with the helix pitch 
$P=150$ nm (at 22$^\circ$C) was used and
the thickness of VADHFLC layer, $d_{FLC}$, 
is maintained at 10~$\mu$m. 
The phase transitions sequence of this LC during heating
up from the solid crystalline phase
is:
Cr$\myarrow{+12}$SmC$^{\star}\myarrow{+110}$SmA$^{\star}\myarrow{+127}$Is,
while during cooling from smectic $C^\star$ phase crystallization occurs
around -10$^\circ$C~ -15$^\circ$C. 
The spontaneous polarization, $P_s$, 
and the tilt angle, $\theta$, at room temperature
are 150~nC/cm$^2$ and 36.5$^\circ$, respectively.

\begin{figure}[!tbh]
\includegraphics[width=8cm]{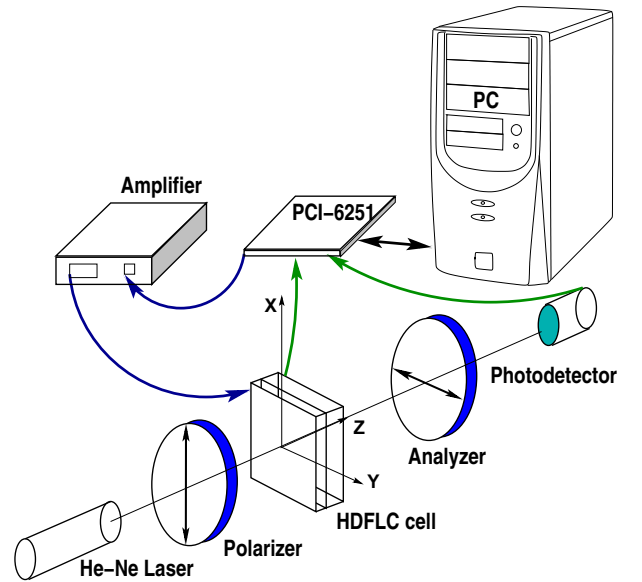}
\caption{%
(Color online)
Experimental setup for electro-optical measurements
in VADHFLC cells. 
}
\label{fig:setup}
\end{figure}

The electro-optical measurements were performed in an automatic
regime. The setup scheme used in the experiments is shown in
Fig.~\ref{fig:setup}. 
The basic element of this experimental setup is computer data acquisition
(DAQ) board NIPCI 6251 from National Instruments.  
A
photo-detector was connected to input board plate for optical measurements.
In our experiments , the output signal, $\pm$10~V, was
not sufficient therefore wideband power amplifier KH model 7600 from
Krohn-Hite Corporation was used. It gives a possibility to have the
output signal up to $\pm$250~V.

\subsection{Results}
\label{subsec:results}

In this section we present and discuss 
the results of our measurements.

\begin{figure}[!tbh]
\centering
\includegraphics[width=9cm]{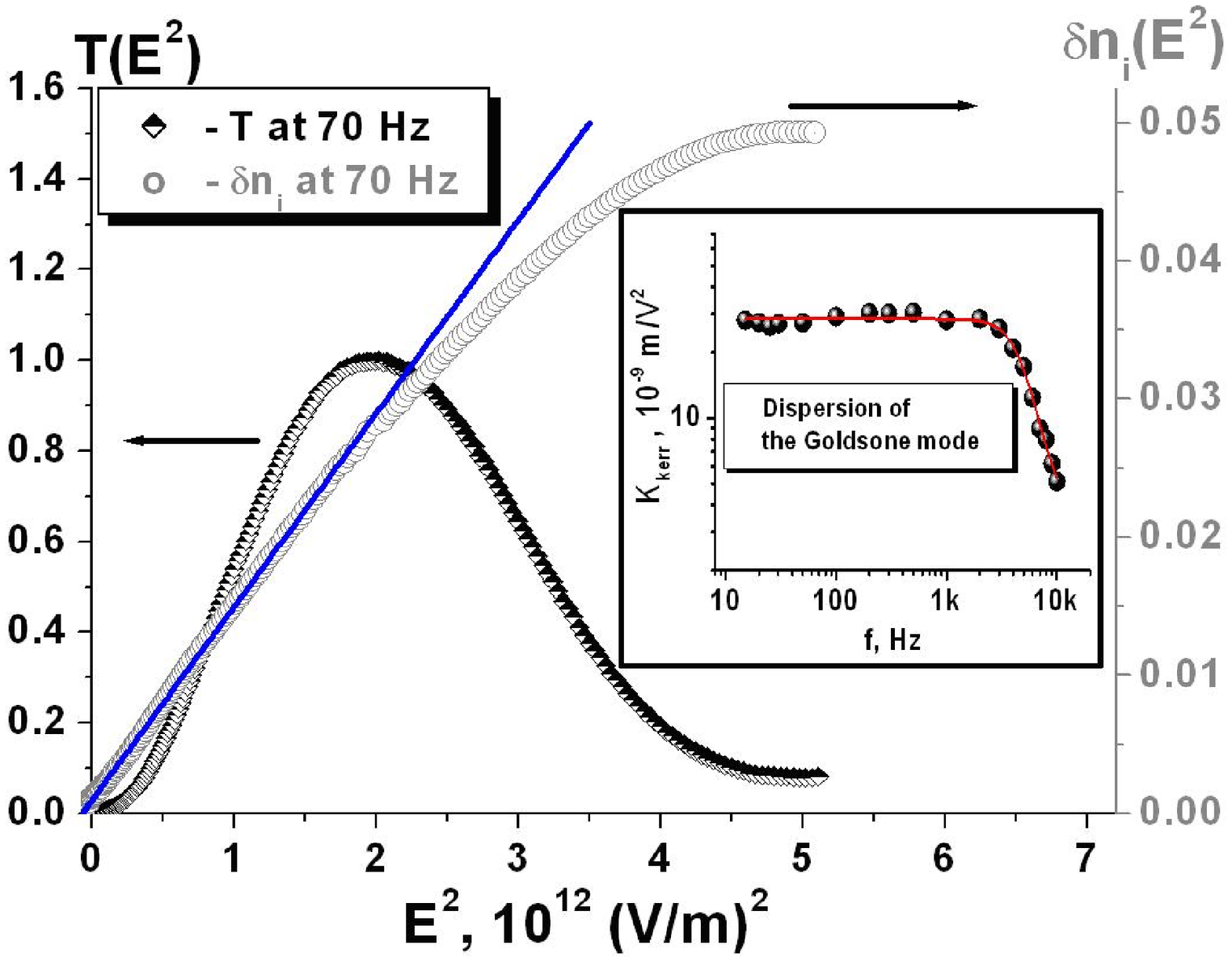}
\caption{%
(Color online)
Electric field dependence of
the transmittance $T(E^2)$ (diamonds) and 
$\delta n_{\ind{ind}}(E^2)$ 
(circles) for VADHFLC
cell placed between crossed polarizers at $\Psi=45^\circ$
(the applied frequency is 70 Hz). 
The solid blue line is the result of theoretical fit 
computed using equation~\eqref{eq:detla_n-vert}. 
Insert shows the
coefficient $K_{\ind{kerr}}$  plotted against frequency. 
}
\label{fig:Transm}
\end{figure}

For a vertically aligned DHFLC cell
placed between crossed polarizers
(see Fig.~\ref{fig:geom}), 
the light transmittance is given by
\begin{align}
  \label{eq:Transm}
  T=\sin^2(2\Psi)\sin^2\frac{\pi\delta n_{\ind{ind}} d_{FLC}}{\lambda},
\end{align}
where 
$d_{FLC}$ is the thickness of FLC layer and
$\Psi$ is
the angle  between the polarization plane of 
incident light and the direction of electric field.
Figure~\ref{fig:Transm} presents
the  experimental data for
the transmittance, $T$, 
measured as a function of $E^2$
at temperature 60$^\circ$C and $\Psi=45^\circ$.  
The corresponding values of electrically induced biaxility,
$\delta n_{\ind{ind}}$, were estimated by using
the formula~\eqref{eq:Transm}
with $\Psi=45^\circ$ and $d_{FLC} = 10$~$\mu$m.
Referring to Fig.~\ref{fig:Transm},
for FLC-587,
the saturation value for 
the $\delta n_{\ind{ind}}$ vs $E^2$ curve 
 is about $0.05$.
The latter is an order of magnitude greater than the value measured 
using the conoscopic method in the smectic $C^\star$ 
phase~\cite{Song:pre:2:2007}
whereas 
it is of the same order of magnitude 
as the values reported
for the smectic $A^\star$
phase~\cite{Bartoli:pre:1997}   
and for the bent core nematic
phase~\cite{Nagaraj:apl:2010}. 

The theoretical curve for
the electric field dependence 
$\delta n_{\ind{ind}} (E^2)$ 
shown in Fig.~\ref{fig:Transm}
as a solid blue line
was calculated from the relation~\eqref{eq:detla_n-vert}
by using the  parameters of
DHFLC measured at temperature 60$^\circ$C: 
$\epsilon_{\perp}=2.3$, $\epsilon_{\parallel}=3.1$, 
$P_s=1.1\times10^{-3}$ C/m$^2$, $\theta= 35^\circ$,
$\chi_E=50$, $P=210$~nm and $\lambda=543$~nm 
(the parameter
$\chi_E$ was measured using an original method published 
in Ref.~\cite{Pozhidaev:mclc:1:2011}, 
$P_s$ was measured by the Sawyer-Tower method~\cite{Sawyer:pr:1930}
).
As it can be seen from Fig.~\ref{fig:Transm},
there are small deviations between 
the theory and the experimental data
when the electric field exceeds 1.5 V/$\mu$m. 
This value is very close to $E_c$
and, therefore, it is reasonable to assume that
the helix unwinding is responsible for such inconsistency. 

With the help of experimental results for $\delta n_{\ind{ind}}$ and
formula~\eqref{eq:detla_n-vert}, 
the coefficient $K_{\ind{kerr}}$ has
been evaluated at different frequencies. 
The resulting frequency dependence of $K_{\ind{kerr}}$ is plotted in 
the insert of Fig.~\ref{fig:Transm}. 
In the low frequency region
where $f<2$~kHz,
the coefficient $K_{\ind{kerr}}$ is constant
and its value $K_{\ind{kerr}}=27$~nm/V$^2$ is about two orders of magnitude
greater than the Kerr constant of nitrobenzene and twice larger than
the Kerr constant of the best known PSBPLC~\cite{Rao:apl:2010}.

At $f>2$~kHz,
the coefficient $K_{\ind{kerr}}$
decreases with the frequency
and
shows a pronounced dispersion.
From equation~\eqref{eq:detla_n-vert}, 
this effect can be attributed to
the dispersion of the dielectric susceptibility of
the Goldstone mode, $\chi_E$,
in the high frequency region.
   
In the static limit, the expression for the Goldstone mode susceptibility
is as follows~\cite{Urbanc:ferro:1991}
\begin{align}
  \label{eq:chi_E}
  \chi_E=\frac{P_s^2}{2 K q^2 \sin^2\theta},
\end{align}
where
$K$ is the elastic constant and
$q=2\pi/P$ is the twist wave number of FLC helix.
This expression can now be combined with
the relation~\eqref{eq:detla_n-vert}
to give the corresponding value of $K_{\ind{kerr}}$.
Hence the Kerr coefficient can be increased further by 
tuning the material parameters
$P_s$, $K$, $q$ and $\theta$ that define the susceptibility
 $\chi_E$.

 As is described in the caption of Fig.~\ref{fig:ellipsoids},
in the presence of electric field parallel to smectic layers,  
the two optical axes  of the VADHFLC cell, 
are rotated through the angle $\psi_d$
(see formula~\eqref{eq:eigenvector} and Fig.~\ref{fig:ellipsoids}) 
in the $x-z$ plane normal to the electric field vector.
In Fig.~\ref{fig:ellipsoids},
this vector is parallel to the $y$ axis and
defines the optical axis which orientation is unaltered. 
It follows that, similar to the well-known B effect in nematic liquid
crystals,
when the polarization vector of incident light is 
perpendicular to the direction of the electric field,
the rotated optical axes remain in the plane of light polarization
and 
the phase modulation of light occurs solely due to 
the change in the effective refractive index.

\begin{figure}[!tbh]
\centering
\includegraphics[width=9cm]{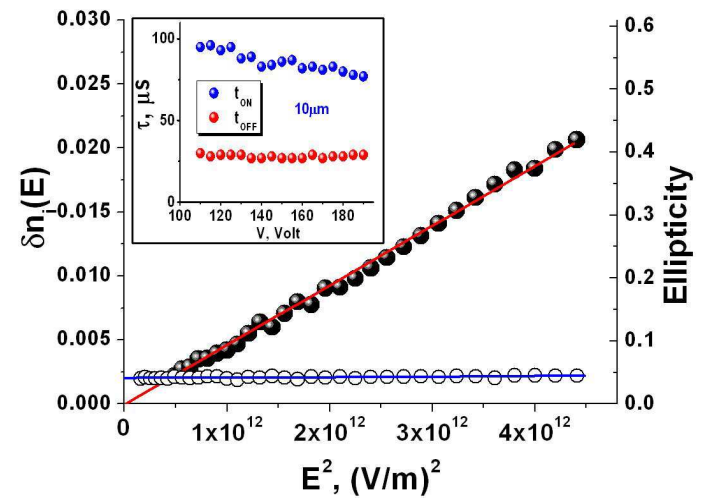}
\caption{%
(Color online)
Solid circles represent $\delta n_{\ind{ind}}$ of 10~$\mu$m thick vertically aligned
DHFLC cell and open circles are for the ellipticity of 
the transmitted light at the frequency of 40Hz.
The ellipticity was measured at $\Psi=90^\circ$. 
Insert shows variation of 
the switching on and switching off times,
$t_{\ind{ON}}$ and $t_{\ind{OFF}}$, with  the applied voltage.
}
\label{fig:ellipt}
\end{figure}

In this case,
the polarization states 
of incident and transmitted light beams
are expected to be identical. 
As it can be seen from Fig.~\ref{fig:ellipt}, no noticeable modulation in the
ellipticity of transmitted light has been observed in our experiments. 
It is clear that, in the low voltage range where
$E<E_c$ and the electric field dependence 
of $\delta n_{\ind{ind}}$ is quadratic, 
the ellipticity is independent of the electric field.
At higher electric fields where $E$ is close to $E_c$,
the quadratic approximation for
the electric field dependence of $\delta n_{\ind{ind}}$
breaks down and 
the ellipticity of the transmitted light begins to grow. 
The helix unwinding process is responsible for 
both of these effects.

Insert in Fig.~\ref{fig:ellipt} represents the electric filed
dependence of 
the switching on and switching off times, $t_{\ind{ON}}$ and $t_{\ind{OFF}}$. 
It is seen that the VADHFLC device is characterized by the fast response
time less than 100~$\mu$s and thus the modulation frequency could be as high
as 2~kHz with saturated electro-optical states. 
Although PSBPLCs
have been recently considered as an emerging display and phase modulating
technology, the VADHFLC mode seems to have the advantage of
higher Kerr constant supplemented with faster
(around an order of magnitude) electro-optical response.
Under certain conditions, EO response of DHFLC cells 
can also be made hysteresis-free~\cite{Pozhidaev:jsid:2012}.  
So, VADHFLCs could find application in many modern electro-optical devices.

\section{Discussion and conclusions}
\label{sec:conclusion}

In this article we have 
studied the electro-optical properties
of VADHFLC cells with subwavelength pitch
that are of key importance for phase modulation.  
Theoretically, it was found that such cells 
can be described by the effective dielectric tensor.
This tensor was evaluated by using an extended version
of the theoretical approach developed in~\cite{Kiselev:pre:2011}.

It was shown that 
the in-plane electric field 
changes the principal values of
the dielectric tensor
so as to transform the  zero-field uniaxial optical anisotropy
into the biaxial one.
In addition, it results in rotation of the
optical axes about the electric field vector.  
It turned out that, in the low voltage limit,
the electrically induced changes of the refractive indices
are characterized by the quadratic nonlinearity
and the above effects can be interpreted as 
the orientational ``Kerr effect''.

Experimentally, we have verified that
the electrically induced biaxiality $\delta n_{\ind{ind}}$ 
shows the quadratic dependence on electric field
in the excellent agreement with 
the predictions of our theoretical model. 
For the noncentrosymmetric systems,
this type of the characteristic is quite uncommon.
In such systems,
the Kerr-like nonlinearity
is typically masked by the dominating contribution from 
the Pockels
electro-optic effect~\cite{Melnichuk:pra:2010}. 
  
The measured Kerr constant $K_{\ind{kerr}}$
for VADHFLC is about 27 nm/V$^2$, which
is about two orders of magnitude larger than that for the nitrobenzene and
twice as large as that for the best known PSBPLC.  
From equations~\eqref{eq:detla_n-vert} and~\eqref{eq:chi_E}
it additionally follows 
that the value of $K_{\ind{kerr}}$ can be increased further by optimizing
the FLC parameters such as the spontaneous polarization and 
the elastic constants with special care of 
the fundamental constraint of $P/\lambda \ll 1$ to 
avoid changes in the polarization characteristics 
such as the ellipticity and the polarization azimuth. 

The 2$\pi$ phase modulation with the
response time less than 100 $\mu$s, 
constant ellipticity of transmitted
light and hysteresis-free electro-optics are the real advantages of
the system. 
This provides an opportunity for the development of a new
generation of phase matrices operating in kilohertz frequency
range. 
Such elements are crucially important for both
science and technology and thus could find application in
many modern electro-optical devices.

 \begin{acknowledgments} 
 This work was supported in part by grants CERG
 612310, CERG 612409 and Russian Foundation of Basic
 Research Grants 13-02-00598\_a,  12-03-90021-Bel\_a.
 \end{acknowledgments} 


%

\end{document}